# The system of uniformly accelerating observers

Bernhard Rothenstein[1)] and Stefan Popescu[2)]
1) Politehnica University of Timisoara, Physics Department,
Timisoara, Romania brothenstein@gmail.com
2) Siemens AG, Erlangen, Germany stefan.popescu@siemens.com

**Abstract.** *We begin with a scenario that involves point-like observers starting at t=0 from the origin O of an inertial reference frame. They move with all possible proper accelerations (0<g<∞) in the positive direction of the OX axis. Equipped with light sources the accelerating observers get involved in experiments like Doppler Effect, radar detection and radar echo. We derive formulas accounting for the experiments mentioned above. In the case of the Doppler Effect we take into account the non-locality aspect in the time interval measurement by accelerating observers.*

## 1. Introduction

The fundamental problem of special relativity can be stated as follows: point-like observers start moving from the same point in space, at the same time and in the same direction with all possible constant speeds (V<c). Find out the relationships between the results they obtain when performing measurements on the same physical object. Changing this scenario, we propose the following problem: **Point-like observers who are able to communicate with each other using light signals start from rest state and from the same point in space and move in the same direction with all possible proper accelerations g (0<g<∞). Find out the relationships between the results they obtain when performing measurements on the same physical object.**

Let $R_0(0,0)$ be a particular observer located at the origin O of its rest inertial reference frame K(XOY). $R'(g)$ is an observer who, relative to the K(XOY) frame, performs a hyperbolic motion characterized by a constant proper acceleration $g$ that begins at a time $t=0$ from the origin O with zero initial velocity. The space coordinate of an event generated by $R'(g)$, attributed by $R_0(0,0)$ is

$$x = \frac{c^2}{g}\left(\sqrt{1+\frac{g^2 t^2}{c^2}} - 1\right) \qquad (1)$$

where $t$ represents the reading of the clock $C_i(x_i,0)$ located at the different points of the OX axis and synchronized following a synchronization procedure proposed by Einstein.[1] The speed of $R'(g)$ relative to K(XOY) at a given time $t$ is

$$V = \frac{gt}{\sqrt{1+\frac{g^2 t^2}{c^2}}}. \qquad (2)$$

Let $C'(g)$ be a clock attached to observer $R'(g)$ which reads $t'=0$ when it is located in front of clock $C_0(0,0)$ attached to the stationary observer $R(0,0)$ which reads $t=0$ as well. It reads $t'$ when it is located in front of a clock $C(x,0)$ of the K(XOY) reference frame that reads $t$. Expressed as a function $t'$ (1) and (2) become

$$x = \frac{c^2}{g}\left(\cosh\frac{g}{c}t' - 1\right) \qquad (3)$$

$$V = c \tanh \frac{g}{c} t' \qquad (4)$$

The two clock readings are related by

$$t = \frac{c}{g} \sinh \frac{g}{c} t' . \qquad (5)$$

**We say that observers $R'_i(g_i)$ moving with different constant proper accelerations $g_i$ in the positive direction of the OX axis of the stationary reference frame K(XOY) make up a system of uniformly accelerating observers.** We should not confound it with the **uniformly accelerating reference frame**[2,3] which involves different uniformly accelerating observers who start with zero initial velocity at $t=0$ from different points of the OX axis located at a distance $x_i = \frac{c^2}{g_i}$ from the origin O of K(XOY). We present in **Figure 1** the world lines $WLR'_1(g_1)$ and $WLR'_2(g_2)$ of two observers $R'_1(g_1)$ and $R'_2(g_2)$ accelerating with $g_1<g_2$ respectively.

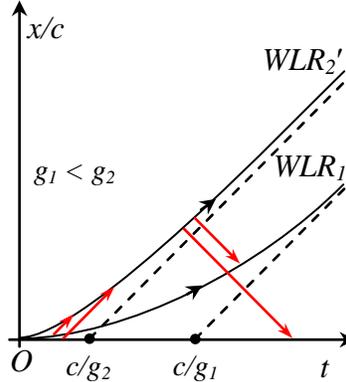

**Figure 1.** *The system of accelerating observers and the possible ways in which they could exchange information via light signals*

We present there the asymptotes of the world lines as well. Equipped with clocks and light sources the accelerating observers get involved with each other in experiments like **Doppler Effect, radar detection and radar echo[4].** They are also involved in the same experiments with a stationary observer $R_0(0,0)$ who handles the light source $S_0(0,0)$ located in front of him. As we see from **Figure 1**:

- An accelerating observer $R'_1(g_1)$ receives all the light signals emitted by the source $S_0(0,0)$ in the time interval $t < c/g_i$, $t$ representing the time displayed by the wrist watch $C_0(0,0)$ of the stationary observer $R_0(0,0)$.
- The accelerating observer $R'_1(g_1)$ receives all the light signals emitted by the accelerating observer $R'_2(g_2)$, $(g_2>g_1)$.
- The accelerating observer $R'_2(g_2)$ receives only the light signals emitted by $R'_1(g_1)$ in the time interval $t < c/g_2$.

The peculiarities of the exchange of information via light signals, between the observers mentioned above, are generated by the fact that after a very long time of motion theirs velocities tend to infinity without reaching it.

## 2. The Doppler Effect in the system of uniformly accelerating observers

### 2.1. Stationary source of light and accelerating receiver

Consider that observer $R_0(0,0)$ emits successive light signals at constant time intervals $T_e$. Let $NT_e$ be the time when the $N^{th}$ light signal is emitted ($N=0,1,2...$) the accelerating observer $R'(g)$ receiving them.

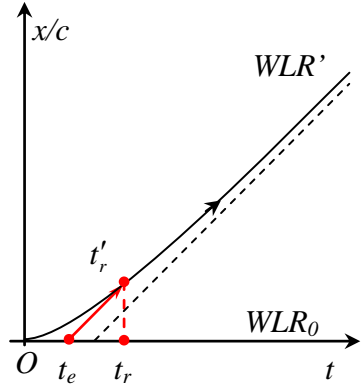

**Figure 2.** *The Doppler experiment with stationary source and an accelerating observer*

The problem is to find out a relationship between the times $t_e = NT_e$ and $t'_{r,N}$ at which the stationary observer emits the successive light signals and at which the accelerating observer receives them respectively. Intersecting the world line of the first observer with the world line of the light signal we have

$$\frac{c^2}{g}\left(\cosh\frac{g}{c}t'_{r,N} - 1\right) = c\left(\frac{c}{g}\sinh\frac{g}{c}t'_{r,N} - NT_e\right). \tag{6}$$

Solved for $t'_{r,N}$ (6) leads to

$$t'_{r,N} = -\frac{c}{g}\ln\left(1 - \frac{gNT_e}{c}\right) \tag{7}$$

where $NT_e$ changes in the range of values $0 < NT_e < c/g_1$. In a similar way we obtain the time $t'_{r,N-1}$ at which $R'(g)$ receives the $(N-1)^{th}$ light signal

$$t'_{r,(N-1)} = -\frac{c}{g}\ln\left[1 - \frac{g(N-1)T_e}{c}\right] \tag{8}$$

If we consider that the observers $R_0(0,0)$ and $R'(g)$ are engaged in a Doppler Effect experiment (stationary source of light $S_0(0,0)$ and uniformly accelerating observer $R'_1(g_1)$) then we characterise it by the Doppler factor

$$D_{0\to 1}(N) = \frac{f_{reception}}{f_{emission}} = \frac{t_{e,N} - t_{e,N-1}}{t'_{r2,N} - t'_{r2,N-1}} = \frac{\frac{g}{c}T_e}{\ln\frac{1-\frac{g}{c}(N-1)T_e}{1-\frac{g}{c}NT_e}} \tag{9}$$

Moreau[5] considers only the light signals emitted at $t=0$ ($N=0$) and at $t=T_e$ ($N=1$) obtaining for the Doppler factor

$$D_M = \frac{T_e}{\frac{c}{g}\ln\frac{1}{1-\frac{g}{c}T_e}} \tag{10}$$

Deriving (9) we have taken into account what physicists call "**non-locality in the period measurement by accelerating observers**"[5] i.e. the fact that the moving observer receives the successive light signals from two different points in space and that during the time interval that separates the two events the velocity of the receiver changes.

Cochran[6] in his approach to this experiment doesn't take into account the non-locality. He starts directly with the formula that accounts for the longitudinal Doppler Effect with light signals

$$D = \sqrt{\frac{1-V/c}{1+V/c}} \tag{11}$$

replacing $V$ with its instantaneous magnitude expressed as a function of $t'_{r,N}$ when the accelerating observer receives the light signal or as a function of the time $NT_e$ when the light signal was emitted. He obtains:

$$D_C(N) = \exp\left(-\frac{g}{c}t'_{r,N}\right) = 1 - \frac{g}{c}NT_e \tag{12}$$

Equations (9) and (12) enable us to evaluate the error committed when ignoring the non-locality effects defined above. Using the shorthand notation $T = c/g$ we obtain:

$$\varepsilon_{[\%]} = \left[1 - \frac{D_C(N)}{D_{0\to 1}(N)}\right]\times 100 = \left[1 - \left(\frac{T}{T_e} - N\right)\ln\frac{\frac{T}{T_e}-(N-1)}{\frac{T}{T_e}-N}\right]\times 100 \tag{13}$$

The frequencies in all the experiments we describe can change in a very large range of values because by modulation we can transmit very small frequencies (acoustic frequencies) but also very large ones (radio frequencies). Low frequencies favour the non-locality whereas the high ones favour the locality. As a consequence we get:

$$\lim_{T_e \to 0} \varepsilon_{[\%]} = 0. \tag{14}$$

We present in **Figure 3** the variation of this error with $T_e/T$ and $N$ as parameters.

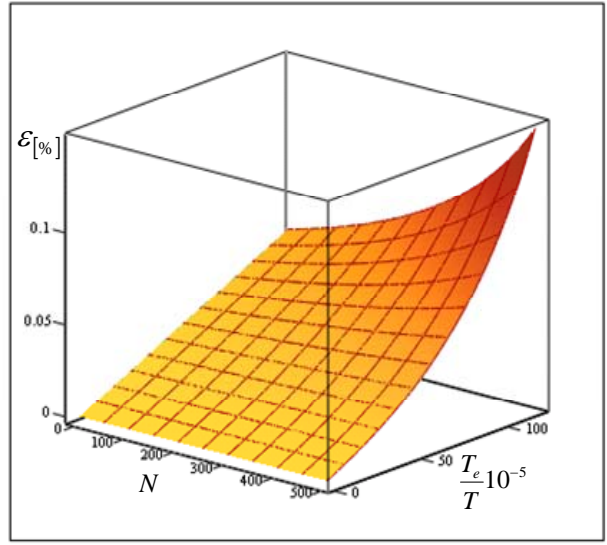

eps

**Figure 3.** *The variation of the error committed by not taking into account the non-locality in the period measurement at very high frequencies*

As this graph reveals, the approximation error decreases with $N$ and $T_e$ and reaches 0 when $T_e$ approaches 0. For $N = 0$ this error is not yet zero but depends on $T_e$.

**2.2. Doppler Effect with stationary receiver and accelerating source of light**

In order to illustrate the asymmetry of the problem under study we consider that the source of light is attached to the accelerating observer $R'(g)$ the stationary observer $R_0(0,0)$ receiving them Figure 4.

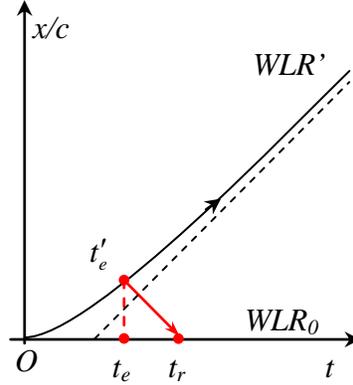

**Figure 4.** *Doppler Effect with stationary receiver and accelerating source*

Intersecting the world line of the accelerating observer with the world line of the light signal it emits we have

$$c\left(t_{r,N} - \frac{c}{g}\sinh\frac{g}{c}NT'_e\right) = \frac{c^2}{g}\left(\cosh\frac{g}{c}NT'_e - 1\right) \qquad (15)$$

where $t_{r,N}$ represents the time when the stationary observer receives the $N^{th}$ light signal emitted by the accelerating source at constant time intervals $T'_e$. Solved for $t_{r,N}$ (15) leads to

$$t_{r,N} = \frac{c}{g}\left(e^{\frac{g}{c}NT'_e} - 1\right). \tag{16}$$

In a similar way we obtain for the $(N-1)^{th}$ emitted light signal

$$t_{r,N-1} = \frac{c}{g}\left(e^{\frac{g}{c}(N-1)T'_e} - 1\right). \tag{17}$$

If we consider that the two observers are involved in a Doppler Effect experiment, we characterize it by a Doppler factor

$$D_{1\to 0}(N) = \frac{f_{reception}}{f_{emission}} = \frac{T'_e}{\frac{c}{g}e^{\frac{g}{c}NT'_e}\left(1 - e^{-\frac{g}{c}T'_e}\right)} \tag{18}$$

where $N$ changes in the range $1 < N < \infty$.

Similar to Cochran's approach we ignore the non-locality by replacing in (11) the non-constant velocity V with its instantaneous value at time when the accelerating source emits the $N^{th}$ light signal. Expressing V as a function of emission time we obtain:

$$D_C(N) = \exp\left(-\frac{g}{c}t'_{e,N}\right) = \exp\left(-\frac{g}{c}NT'_e\right) \tag{19}$$

With (19) the error introduced by ignoring the non-locality becomes:

$$\varepsilon_{[\%]} = \left[1 - \frac{D_C(N)}{D_{1\to 0}(N)}\right] \times 100 = \left[1 - \frac{T'_e}{T} \frac{e^{-\frac{NT'_e}{T}}}{e^{\frac{NT'_e}{T}}\left(1 - e^{-\frac{T'_e}{T}}\right)}\right] \times 100 \tag{20}$$

and we present its variation in Figure 5.

The approximation error decreases with $N$ and $T_e$ and reaches 0 when $T_e$ approaches 0. For $N = 0$ this error is not yet zero but depends on $T_e$. It is also interesting to notice that the maximum error is limited to 100% in this case.

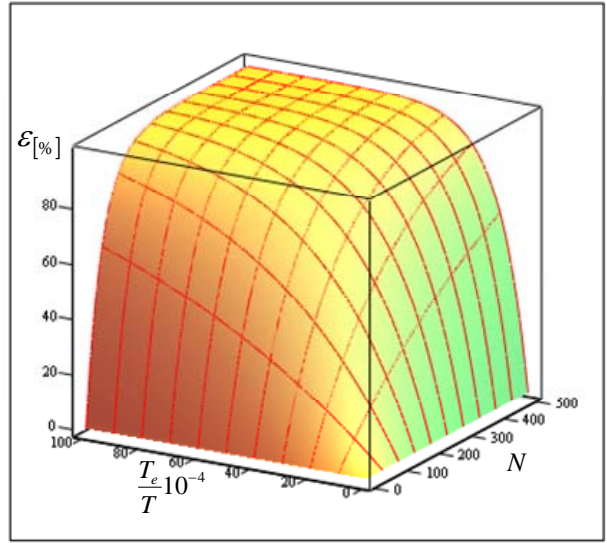

**Figure 5.** *The variation of the error introduced by not taking into account the non-locality in the period measurement at very high frequencies*

## 2.3 Source of light co-moving with $R'_1(g_1)$ and $R'_2(g_2)$ as receiver (Figure 6)

The problem is to find out a relationship between the constant period $T'_{e1}$ at which the source emits successive light signals and the time interval $T'_{r2}$ during which the accelerating observer receives them.

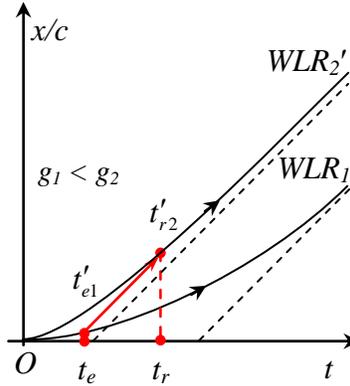

**Figure 6.** *The Doppler experiment with two accelerating observers. Observer $\mathbf{R'_1}$ emits the light signals $\mathbf{R'_2}$ receiving them*

In this experiment we have $t_e < c/g_2$ and in the equations derived below we have $T'_{e,1} < c/g_1$. We begin by considering the emission of a single light signal from observer $R'_1(g_1)$ at proper time $t'_{e2}$ and its reception at receiver $R'_1(g_1)$ at proper time $t'_{r1}$. We are looking for to find the relation between these proper times. In the world diagram above we have in terms of the stationary rest frame K(XOY) the following relation:

$$c(t_r - t_e) = x_r - x_e \qquad (21)$$

By grouping the terms for the same event and transposing them in proper time coordinates of the respective observers we obtain:

$$\frac{1}{g_1}\left[\cosh\left(\frac{g_1}{c}t'_{e1}\right)-1-\sinh\left(\frac{g_1}{c}t'_{e1}\right)\right] = \frac{1}{g_2}\left[\cosh\left(\frac{g_2}{c}t'_{r2}\right)-1-\sinh\left(\frac{g_2}{c}t'_{r2}\right)\right] \qquad (22)$$

and after expanding the hyperbolic terms we get the simplified relation:

$$\frac{1}{g_1}\left(e^{-\frac{g_1}{c}t'_{e1}}-1\right) = \frac{1}{g_2}\left(e^{-\frac{g_2}{c}t'_{r2}}-1\right) \qquad (23)$$

Consider now that observer $R'_1(g_1)$ emits periodic signals at regular proper time instants $t'_{e1,N} = N \cdot T'_{e1}$. Observer $R'_2(g_2)$ will receive these signals at proper reception times:

$$t'_{r2,N} = -\frac{c}{g_2}\ln\left[1+\frac{g_2}{g_1}\left(e^{-\frac{g_1}{c}t'_{e1,N}}-1\right)\right] \qquad (24)$$

This last equation allows us to calculate the Doppler factor that characterizes this experiment as:

$$D_{1\to 2}(N) = \frac{t'_{e1,N}-t'_{e1,N-1}}{t'_{r2,N}-t'_{r2,N-1}} = \frac{g_2}{g_1}\frac{\frac{g_1}{c}T'_{e1}}{\ln\frac{1+\frac{g_2}{g_1}\left(e^{-\frac{g_1}{c}(N-1)T'_{e1}}-1\right)}{1+\frac{g_2}{g_1}\left(e^{-\frac{g_1}{c}NT'_{e1}}-1\right)}} \qquad (25)$$

The graphical representation below (Figure 7) shows that this Doppler factor decreases with the order number *N* and the relative acceleration ratio *g₂/g₁*. This behaviour can be easily deduced from figure 6 were we notice that, because of the hyperbolic motion, even for shorter emission periods the reception periods rapidly increases with *N*.

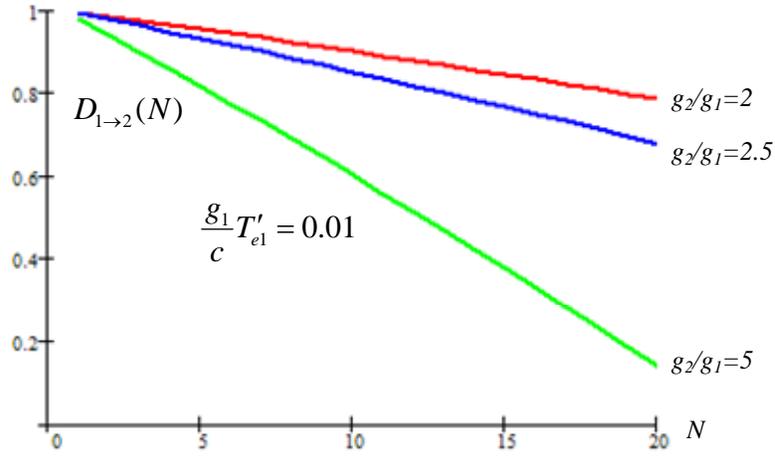

**Figure 7.** *The variation of Doppler factor with the order number N*

### 2.4 Source of light commoving with $R'_2(g_2)$ and $R'_1(g_1)$ as receiver (Figure 8)

This reversed scenario is depicted in the figure 7 below. Now we are looking for to find the new relation between the proper emission and reception times.

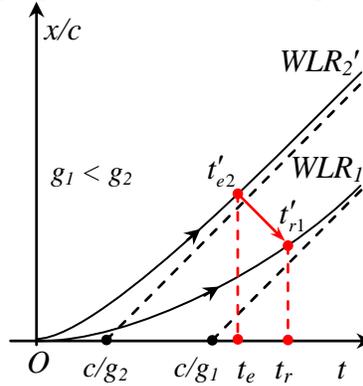

**Figure 8.** *The Doppler experiment with two accelerating observers. Observer $\mathbf{R'_2}$ emits the signals.*

In the world diagram above we have in terms of the stationary rest frame K(XOY) the following relation:

$$c(t_r - t_e) = x_e - x_r \tag{26}$$

By following the same calculation method as above we find the new relation between the proper emission and reception times as:

$$\frac{1}{g_2}\left(e^{\frac{g_2 t'_{e2}}{c}} - 1\right) = \frac{1}{g_1}\left(e^{\frac{g_1 t'_{r1}}{c}} - 1\right) \tag{27}$$

Again as above the last equation allows us to calculate the Doppler factor that characterizes this experiment as:

$$D_{2\to1}(N) = \frac{t'_{e2,N} - t'_{e2,N-1}}{t'_{r1,N} - t'_{r1,N-1}} = \frac{g_1}{g_2} \frac{\frac{g_2}{c}T'_{e2}}{\ln\frac{1+\frac{g_1}{g_2}\left(e^{\frac{g_2}{c}NT'_{e2}} - 1\right)}{1+\frac{g_1}{g_2}\left(e^{\frac{g_2}{c}(N-1)T'_{e2}} - 1\right)}} \qquad (28)$$

The graphical representation of the Doppler factor in this case reveals that it quickly converges toward the final value $g_1/g_2$ as $N$ increases (Figure 9). A closer look at Figure 8 allows for a easier understanding of this effect. For large emission times the world lines of the two observers become parallel and thus the difference between the emission and reception periods remains about the same. As expected we reach this convergence faster with longer emission periods.

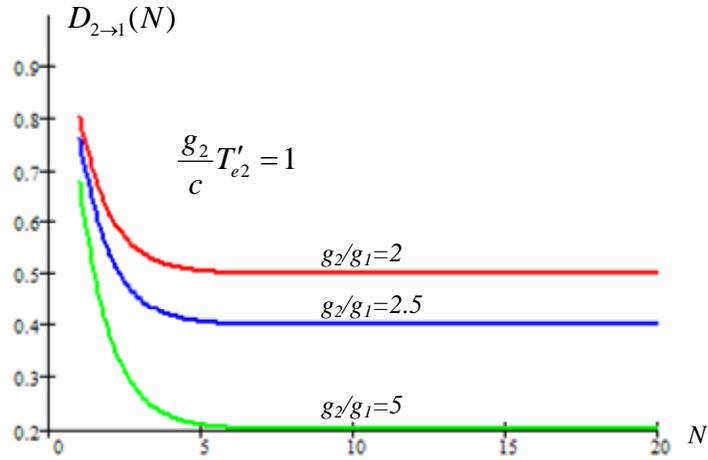

**Figure 9.** *The variation of Doppler factor with the order number N*

## 3. Radar detection

### 3.1. Observer $R'_1(g_1)$ detects the space-time coordinates of an event that takes place on the world line of observer $R'_2(g_2)$ using the radar detection method[4]

In the experiment we propose, illustrated in Figure 10, observer $R'_1(g_1)$ detects the space-time coordinates of an event that takes place on the world line of observer $R'_2(g_2)$ using the radar detection method. The following events are involved in this experiment:

- **Emission - $E_e(x_e, t_e)$** → Observer $R'_1(g_1)$ at position $x_e$ and time $t_e$ (proper time $t'_{e1}$) emits a light signal towards observer $R'_2(g_2)$;
- **Mirroring - $E_m(x_m, t_m)$** → Observer $R'_2(g_2)$ receives the light signal at position $x_m$ and time $t_m$ (proper time $t'_{m2}$) and reflects it back without delay towards observer $R'_1(g_1)$;
- **Reception - $E_r(x_r, t_r)$** → Observer $R'_1(g_1)$ receives the reflected light signal at position $x_r$ and time $t_r$ (proper time $t'_{r1}$) and calculates the radar detected coordinates of $R'_2(g_2)$.

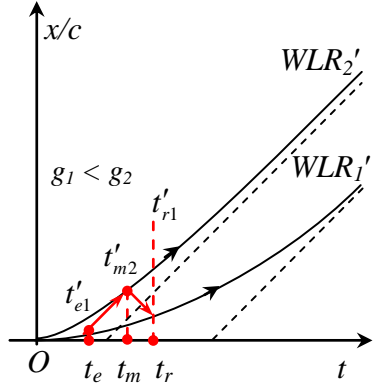

**Figure 10.** *The radar experiment with two accelerating observers. Observer $R'_1$ performs the experiment.*

This radar experiment is only possible for emission times $t_e < c/g_2$. The problem now is to express the space-time coordinates of event $E_m(x_m, t_m)$ that takes place on the world line of $R'_2(g_2)$ as a function of the times $t'_{e,1}$ and $t'_{r,1}$ displayed by the wrist watch of observer $R'_1(g_1)$. In the inertial frame at rest we have the following obvious relations:

$$c(t_m - t_e) = x_m - x_e$$
$$c(t_r - t_m) = x_m - x_r \quad (30)$$

This radar experiment is only possible for emission times $-\infty < t_e < c/g_2$. Rearranging the equations above for the respective events we obtain:

$$x_e - ct_e = x_m - ct_m$$
$$x_r + ct_r = x_m + ct_m.$$
(31)

After transposing in proper time coordinates of the respective observers

$$\frac{1}{g_1}\left[\cosh\left(\frac{g_1}{c}t'_{e1}\right) - 1 - \sinh\left(\frac{g_1}{c}t'_{e1}\right)\right] = \frac{1}{g_2}\left[\cosh\left(\frac{g_2}{c}t'_{m2}\right) - 1 - \sinh\left(\frac{g_2}{c}t'_{m2}\right)\right]$$

$$\frac{1}{g_1}\left[\cosh\left(\frac{g_1}{c}t'_{r1}\right) - 1 + \sinh\left(\frac{g_1}{c}t'_{r1}\right)\right] = \frac{1}{g_2}\left[\cosh\left(\frac{g_2}{c}t'_{m2}\right) - 1 + \sinh\left(\frac{g_2}{c}t'_{m2}\right)\right]$$
(32)

and expanding the hyperbolic terms we get the simplified relations:

$$\frac{1}{g_1}\left(e^{-\frac{g_1}{c}t'_{e1}} - 1\right) = \frac{1}{g_2}\left(e^{-\frac{g_2}{c}t'_{m2}} - 1\right)$$

$$\frac{1}{g_1}\left(e^{\frac{g_1}{c}t'_{r1}} - 1\right) = \frac{1}{g_2}\left(e^{\frac{g_2}{c}t'_{m2}} - 1\right)$$
(33)

Eliminating $t'_{2,m}$ above we obtain that the emission and reception times are related by

$$t'_{r,1} = \frac{c}{g_1}\ln\frac{2 - \frac{g_2}{g_1} - \left(1 - \frac{g_2}{g_1}\right)e^{-\frac{g_1}{c}t'_{e,1}}}{1 + \frac{g_2}{g_1}\left(e^{-\frac{g_1}{c}t'_{e,1}} - 1\right)}.$$
(34)

By definition, observer $R'_1(g_1)$ assigns to the reflection event taking place on the world line of observer $R'_2(g_2)$ a radar detected distance to target $X'_{1\to 2,radar} = \frac{c}{2}(t'_{r,1} - t'_{e,1})$ and a radar detected target time coordinate $t'_{1\to 2,radar} = \frac{1}{2}(t'_{r,1} + t'_{e,1})$. Expressing these space-time coordinate as a function of the proper time of the active performing observer $R'_1(g_1)$ we obtain:

$$X'_{1\to 2,radar}(t'_{e,1}) = \frac{c}{2}\left[\frac{c}{g_1}\ln\frac{2 - \frac{g_2}{g_1} - \left(1 - \frac{g_2}{g_1}\right)e^{-\frac{g_1}{c}t'_{e,1}}}{1 + \frac{g_2}{g_1}\left(e^{-\frac{g_1}{c}t'_{e,1}} - 1\right)} - t'_{e,1}\right]$$
(35)

and a radar detected target time coordinate:

$$t'_{1\to 2,radar}(t'_{e,1}) = \frac{1}{2}\left[\frac{c}{g_1}\ln\frac{2-\frac{g_2}{g_1}-\left(1-\frac{g_2}{g_1}\right)e^{-\frac{g_1}{c}t'_{e,1}}}{1+\frac{g_2}{g_1}\left(e^{-\frac{g_1}{c}t'_{e,1}}-1\right)} + t'_{e,1}\right] \qquad (36)$$

As detected by $R'_1(g_1)$ the radar distance between observers is time dependent and the relationship between the two times is non-linear. If $g_1 = g_2$ then the distance occurs to be always zero as the two accelerating observes permanently superpose their positions whilst the time readings coincide as the two accelerating observes are always situated at the same point in space. Equation (35) and (36) are the parametric equations for the motion of $R'_2(g_2)$ as detected by observer $R'_1(g_1)$. We depict figure (11) the world line of $R'_2(g_2)$ represented in a <u>radar world diagram</u> as perceived by observer $R'_1(g_1)$ who performs the radar detection. This appears to approximate the hyperbolic motion.

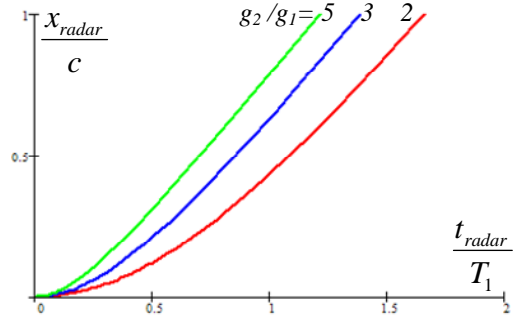

**Figure 13.** *The world line of observer* **R$'_2$** *in the radar word diagram of observer* **R$'_1$**.

## 3.2. Observer $R'_2(g_2)$ detects the space-time coordinates of an event that takes place on the world line of observer $R'_2(g_2)$ using the radar detection method

Lets now reverse the scenario above and suppose that observer $R'_2(g_2)$ is performing the active role in the radar detection experiment. The scenario we follow is sketched in Figure 14 involving the following events:

- **Emission - E$_e$**($x_e,t_e$) → Observer $R'_2(g_2)$ at position $x_e$ and time $t_e$ (proper time $t'_{e2}$) emits a light signal towards observer $R'_1(g_1)$;
- **Mirroring - E$_m$**($x_m,t_m$) → Observer $R'_1(g_1)$ receives the light signal at position $x_m$ and time $t_m$ (proper time $t'_{m1}$) and reflects it back without delay towards observer $R'_2(g_2)$;
- **Reception - E$_r$**($x_r,t_r$) → Observer $R'_2(g_2)$ receives the reflected light signal at position $x_r$ and time $t_r$ (proper time $t'_{r2}$) and calculates the radar detected coordinates of $R'_1(g_1)$.

**Figure 14.** *The radar experiment with two accelerating observers. Observer $R'_2$ performs the experiment.*

This radar experiment is only possible for those emission times $t_e$ leading to mirroring times $t_m < c/g_2$. In the inertial frame at rest we have the following obvious relations:

$$c(t_m - t_e) = x_e - x_m$$
$$c(t_r - t_m) = x_r - x_m \qquad (37)$$

Following the same calculation method as above we obtain that the emission and reception times are related by

$$t'_{r,1} = \frac{c}{g_2} \ln \frac{1 + \frac{g_1}{g_2}\left(e^{\frac{g_2}{c}t'_{e,2}} - 1\right)}{2 - \frac{g_1}{g_2} - \left(1 - \frac{g_1}{g_2}\right)e^{\frac{g_2}{c}t'_{e,2}}} \qquad (38)$$

Again by definition, observer $R'_2(g_2)$ assigns to the reflection event taking place on the world line of observer $R'_1(g_1)$ a radar detected distance to target:

$$X'_{2\to 1, radar}(t'_{e,2}) = \frac{c}{2}\left[\frac{c}{g_2}\ln\frac{1 + \frac{g_1}{g_2}\left(e^{\frac{g_2}{c}t'_{e,2}} - 1\right)}{2 - \frac{g_1}{g_2} - \left(1 - \frac{g_1}{g_2}\right)e^{\frac{g_2}{c}t'_{e,2}}} - t'_{e,2}\right] \qquad (39)$$

and a radar detected target time coordinate:

$$t'_{2\to 1, radar}(t'_{e,2}) = \frac{1}{2}\left[\frac{c}{g_2}\ln\frac{1 + \frac{g_1}{g_2}\left(e^{\frac{g_2}{c}t'_{e,2}} - 1\right)}{2 - \frac{g_1}{g_2} - \left(1 - \frac{g_1}{g_2}\right)e^{\frac{g_2}{c}t'_{e,2}}} + t'_{e,2}\right] \qquad (40)$$

Again we interpret (39) and (40) as being the parametric equations for the world line of observer $R'_1(g_1)$ as represented in the "radar world coordinates" of observer $R'_2(g_2)$. The resulted trajectory appears to approximate the hyperbolic motion again.

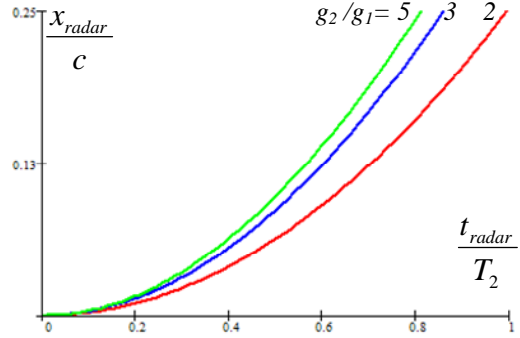

**Figure 15.** *The world line of observer $\mathbf{R}'_1$ in the radar word diagram of observer $\mathbf{R}'_2$.*

### 4. The radar echo with accelerating observers

**4.1. Observer $\mathbf{R}'_1(\mathbf{g}_1)$ emits light signals at constant time intervals and receives them back after reflection on a mirror co-moving with observer $\mathbf{R}'_2(\mathbf{g}_2)$.**

The scenario followed in the case of the radar detection and the results obtained in **3.1** enable us to solve the **radar echo** problem: Observer $R'_1(g_1)$ emits successive light signals, at constant time intervals, towards observer $R'_2(g_2)$ and receives them back after theirs reflection on a mirror co-moving with the second observer. The problem is to find out a relationship between the constant emission periods $\Delta t'_{e,1}$ and the reception periods $\Delta t'_{r,1}$. We start with the relationship (34) that relates the emission and reception times. Differentiating both sides we obtain the formula that accounts for the radar echo as:

$$\Delta t'_{r1} = \frac{e^{-\frac{g_1}{c}t'_{e1}}}{\left[2 - \frac{g_2}{g_1} + \left(\frac{g_2}{g_1} - 1\right)e^{-\frac{g_1}{c}t'_{e1}}\right] \cdot \left[1 - \frac{g_2}{g_1} + \frac{g_2}{g_1}e^{-\frac{g_1}{c}t'_{e1}}\right]} \Delta t'_{e1} \qquad (41)$$

Introducing the relative acceleration constant defined as $k_g = \frac{g_2}{g_1} - 1 \geq 0$, we obtain the instantaneous radar Doppler factor as:

$$D_{1 \to 2, radar} = \frac{\Delta t'_{e1}}{\Delta t'_{r1}} = \left[1 - k_g + k_g e^{-\frac{g_1}{c}t'_{e1}}\right] \cdot \left[1 + k_g - k_g e^{\frac{g_1}{c}t'_{e1}}\right] \qquad (42)$$

with $D_{1\to2,radar} \to 1$ when $k_g \to 0$.

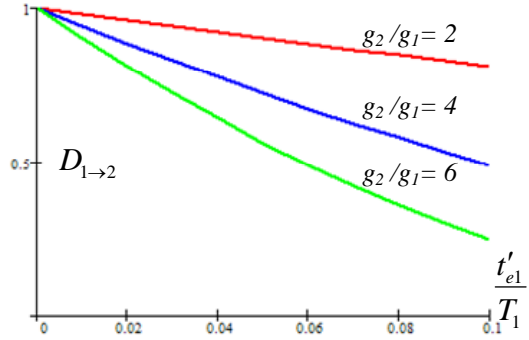

**Figure 16.** *The radar echo Doppler factor when observer $\mathbf{R}'_1$ performs the experiment.*

### 4.2 Observer $\mathbf{R}'_2(g_2)$ emits light signals at constant time intervals and receives them back after reflection on a mirror co-moving with observer $\mathbf{R}'(g_1)$

The scenario followed in the case of the radar detection and the relationship between the emission and the reception times (38) leads by differentiation to the following relationship between the emission period and reception period that accounts for the radar echo:

$$\Delta t'_{r2} = \frac{e^{\frac{g_2}{c}t'_{e2}}}{\left[2 - \frac{g_1}{g_2} + \left(\frac{g_1}{g_2} - 1\right)e^{\frac{g_2}{c}t'_{e2}}\right] \cdot \left[1 - \frac{g_1}{g_2} + \frac{g_1}{g_2}e^{\frac{g_2}{c}t'_{e2}}\right]} \Delta t'_{e2} \qquad (43)$$

Again introducing the complementary relative acceleration constant, this time defined as $k_g^* = \frac{g_1}{g_2} - 1$ with $-1 \le k_g^* \le 0$, we obtain the instantaneous radar Doppler factor as:

$$D_{2\to1,radar} = \frac{\Delta t'_{e2}}{\Delta t'_{r2}} = \left[1 - k_g^* + k_g^* e^{\frac{g_2}{c}t'_{e2}}\right] \cdot \left[1 + k_g^* - k_g^* e^{-\frac{g_2}{c}t'_{e2}}\right] \qquad (44)$$

with $D_{2\to1,radar} \to 1$ when $k_g^* \to 0$.

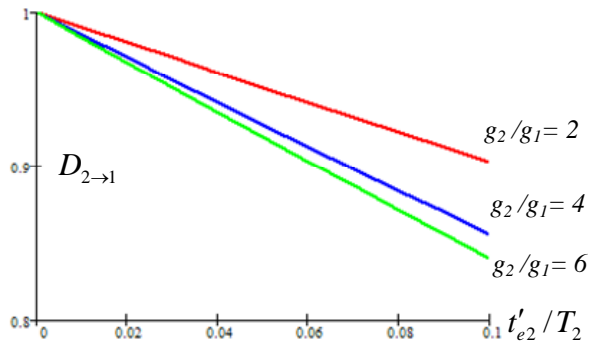

**Figure 17.** *The radar echo Doppler factor when observer $\mathbf{R}'_2$ performs the experiment.*

The equations that account for the radar echo derived above hold only in the case of very high frequencies (locality assumption). It is very interesting to notice in figures 14 and 15 that the Doppler factor for both radar echo scenarios analysed above is always smaller than 1. It means that the reception period is always longer than the emission period, independent on which observer accelerates faster or performs the radar experiment. This effect may be better understood by a closer look at figures 6 and 8. It appears very clear that because of the hyperbolic motion we witness a significant period dilation effect accumulated during the time when the light signal proceeds from observer $\mathbf{R}'_1$ toward observer $\mathbf{R}'_2$.

## 5. Conclusions

We considered a scenario involving observers performing the hyperbolic motion with all possible proper accelerations $0 < g < \infty$. This enabled us to reveal the peculiarities of the experiments these observers perform involving information exchange via light signals they emit and receive. The problem has not only an academic interest, the results obtained being of interest also for the pilots of high-speed airplanes.